\documentclass[twocolumn]{article} 
\usepackage{graphicx, amsmath, amssymb, amsfonts, mathrsfs}
\usepackage[latin1]{inputenc}
\usepackage[T1]{fontenc} 
\usepackage{textcomp}
\title{Nuclear spin interferences in bulk water at room temperature.}
\author{J.~Grucker\footnote{Laboratoire de Physique des Lasers - UMR7538 CNRS-Université Paris 13. 93430 VILLETANEUSE, FRANCE} \and K.~van~Schenk~Brill\footnote{Laboratoire de Neuroimagerie in Vivo - UMR7004 ULP-CNRS.  67085 STRASBOURG Cedex, FRANCE} 
\and E.~Belaga\footnote{Institut de Recherche en Mathématiques Avancées - UMR7501 ULP-CNRS. 67084 STRASBOURG, France} \and J.~Baudon\footnotemark[1] \and D.~Grucker \footnotemark[2] \thanks{Corresponding author: grucker@ipb.u-strasbg.fr}}            
\begin{document}
\maketitle
\noindent \textsc{pacs} 82.56.Jn: Pulse sequences in NMR\\
\textsc{pacs} 03.67.-a: Quantum information\\
\textsc{pacs} 67.57.Lm: Spin dynamics\\
\begin{abstract}
Nuclear spin interference effects generated in a macroscopic sample of $10\ \mathrm{ml}$ degassed water are detected in a simple NMR experiment. A $\pi/2\!-\!\tau\!-\!\pi/2$~RF double pulse sequence (Ramsey sequence) is applied to the water sample immersed in a static magnetic field $B_0~\!\approx\!~4.7\ \mathrm{T}$. For a homogeneity of $B_0$ of the order of $\Delta B_0/B_0=2\cdot 10^{-8}$, the nuclear spin interference term is controlled with a maximum relative deviation of  $9.7~\%$. These results are a first step to manipulation of nuclear spin coherence of water molecules.
\end{abstract}
\section{Introduction}
Recent developments of quantum computing have emphasized the importance of quantum interferences for quantum information processing~(QIP)~\cite{qcqi}. The best known experimental quantum computing demonstration was obtained by bulk Nuclear Magnetic Resonance~(NMR) of perfluorobutadienyl molecule corresponding to seven qubits~\cite{7q}. A similar, more realistic and self-contained experimental NMR realization, using  tert-butoxycarbonyl-($^{13}\!\mathrm{C}_{2}\!-^{15}\!\mathrm{N}\!-^{2}\!~\mathrm{D}_{2}^\alpha$-glycine) fluoride molecules, produced a quantum computer on five qubits~\cite{5q}.  In  NMR QIP, a qubit is defined by the resonance frequency of a nuclear spin in its local magnetic field i.e. the applied magnetic field~$B_0$ corrected by a shield due to the surrounding electrons. The initial step of QIP is to define a pure quantum state which is masked by the fact that for bulk~NMR the spin system is a statistical ensemble and the correct description of the system is given by the use of the density matrix formalism. The $J$~coupling between two nuclear spins induces coherences of nuclear spin states revealing, via interference effects, the quantum nature of  the system. The first demonstration of spinor character for spin~$\frac{1}{2}$ nucleus via NMR~interferometry  using the $J$~
coupling was performed by M.~E.~Stoll et al. using $^{13}\!\mathrm{C}$ enriched sodium formate ($\mathrm{NaCHO}_2$) dissolved in $\mathrm{D}_2\mathrm{O}$~\cite{spinstoll}. In absence of a constant coupling  between the spin states,  spin interferences have been observed with a neutron beam interferometer in 1975~\cite{neutr}. Already in 1950, N.~F.~Ramsey showed that two successive $\pi/2$~pulses can induce atomic state interferences, the so called Ramsey fringes~\cite{ramsey}. Recently it was shown that superconducting tunnel junction circuit displayed signatures of macroscopic quantum behavior as Ramsey fringes~\cite{scond}. Here we will show that such interference effects can be detected in bulk water at room temperature.
\section{Theoretical background}
The theoretical description of the system considers a macroscopic amount of $N$ identical spins $I=\frac{1}{2}$ immersed in a static $B_0$~magnetic field for a long time. The Hamiltonian of this system is ($\hbar=1$):
\begin{equation}
\label{Hamil}
\mathscr{H}=-\mu\cdot B_0=\gamma\cdot I\cdot B_0,
\end{equation}
where $\gamma$ is the gyromagnetic ratio of the spin~$I$. By convention, $B_0$ gives the $z$~direction of the coordinates and the $x0y$~plane is called the transverse plane. The magnetic field direction~$z$ is the most natural quantum axis along which one can define two eigenvectors $|-\rangle$ and $|+\rangle$ of respective energies:
\begin{eqnarray}
    \label{eigenenergy}
     E_{-}&=&-\frac{1}{2}\gamma\cdot B_0=-\frac{\omega_0}{2} \nonumber \\
     E_{+}&=&\frac{1}{2}\gamma\cdot B_0=\frac{\omega_0}{2}, 
\end{eqnarray}
where $\omega_0 =  \gamma\cdot B_0$ is the angular frequency of the allowed transition between $|-\rangle$ and $|+\rangle$. As there is a statistical ensemble of $N$ nuclear spins, one has to consider the density matrix of the system. At thermal equilibrium, the population of each eigenstate is given by the Boltzmann distribution and then the density matrix can be written as: 
\begin{equation}
    \rho_0=\frac{1}{2}\left(\left(1-d\right)|+\rangle\langle+|+
\left(1+d\right)|-\rangle\langle-|\right), \nonumber
\end{equation}
where one made the approximation $e^{\pm\omega_0/2kT}\approx 1\pm d$, with $d=\omega_0/2kT$. The NMR signal originates from transverse magnetic moments $M_x$ and~$M_y$. For matter of clarity, one can only consider~$M_x$ which is given by the trace of the $\rho\cdot S_x$~matrix (noted $M_x=\mathrm{Tr}\left[\rho\cdot S_x\right]$) where $S_x  =\frac{1}{2}\left(|+\rangle\langle-|+|-\rangle\langle+|\right)$~\cite{tannou}. 
One can notice that $\mathrm{Tr}\left[\rho_0\cdot S_x\right]=0$, meaning that there is no NMR signal at thermal equilibrium as expected~\cite{abra,slic}. The system can be modified by a perturbation obtained by a polarized RF~wave generated by an oscillating current in a resonator at the frequency $\omega_0/2\pi$. An RF~pulse acts on a nuclear spin as a rotation of Euler angles $(\theta,\beta)$ in the spin space~\cite{tannou}. In the $\left(|-\rangle,|+\rangle\right)$ basis, the rotation matrix for spin~$\frac{1}{2}$ is:
\begin{eqnarray}
 r[\theta,\beta]&=&\cos\frac{\theta}{2}e^{\frac{-i\beta}{2}}|+\rangle\langle+|+
        \sin\frac{\theta}{2}e^{\frac{i\beta}{2}}|+\rangle\langle-|\nonumber\\
   &&-\sin\frac{\theta}{2}e^{\frac{-i\beta}{2}}|-\rangle\langle+|+
\cos\frac{\theta}{2}e^{\frac{i\beta}{2}}|-\rangle\langle-|. \nonumber   
\end{eqnarray}
In term of the density matrix, a $(\theta,\beta)$~rotation of~$\rho_0$ gives a rotated $\rho_1$ density matrix according to:  $\rho_1=r^{-1}[\theta,\beta]\cdot  \rho_0\cdot r[\theta,\beta]$, where $r^{-1}[\theta,\beta]\cdot r[\theta,\beta]=1$ (unity matrix). We first take $\beta=0$. After a single pulse~$\theta$, $M_x[\theta]$ is given by: $M_x[\theta]=\mathrm{Tr}\left[r^{\dagger}\left[\theta,0\right]\cdot \rho_0\cdot r\left[\theta,0\right]\cdot S_x\right] = d/2 \sin\theta$. As it is well known, one finds that after a single RF~pulse the signal is maximum for $\theta = \pi/2$ (the RF~pulse is called a $\pi/2$~pulse) and zero for $\theta = \pi $ ($\pi$~pulse). The issue of the experiment described in this article is to produce a sequence of two consecutive $\pi/2$~pulses separated by a time~$\tau$ during which the eigenstates follow their own free evolution. It is the equivalent for nuclear spins in water molecules to the famous N.~Ramsey experiment for a beam of potassium atoms~\cite{ramsey}. The sample is first shined by a $\pi/2$~pulse, which means that the initial density matrix~$\rho_0$ turns to be:
\begin{eqnarray}
    \rho_1&=&r^{\dagger}[\pi/2,0]\rho_0 r[\pi/2,0]\nonumber\\
          &=&\frac{1}{2}|+\rangle\langle+|+
            \frac{d}{2}|+\rangle\langle-|+\frac{d}{2}|-\rangle\langle+|
           +\frac{1}{2}|-\rangle\langle-|\nonumber.
\end{eqnarray}
During a time~$\tau$, the density matrix~$\rho_1$ is free to evolve in $B_0$.  This free evolution obeys to the equation $i \frac{\mathrm{d} \rho_1}{\mathrm{d} t} = \mathscr{H}\rho_1-\rho_1\mathscr{H}$, where $\mathscr{H}$ is given by equation~(\ref{Hamil}). One can easily show that after time~$\tau$, the density matrix of the system turns to be:
\begin{eqnarray}
    \rho_2 &=& \frac{1}{2}|+\rangle\langle+| 
              + \frac{d}{2}e^{-i\omega_0\tau}|+\rangle\langle-| \nonumber \\
           && +\frac{d}{2}e^{i\omega_0\tau}|-\rangle\langle+|
              +\frac{1}{2}|-\rangle\langle-|\nonumber.
\end{eqnarray}
If one then applies a second $r[\pi/2,\beta]$~rotation to~$\rho_2$, the final density matrix~$\rho_3$ after this $\pi/2\!-\!\tau\!-\!\pi/2$~sequence is $\rho_3=r^{\dagger}[\pi/2,\beta]\rho_0 r[\pi/2,\beta]$, where $\beta$~is the relative phase of the two $\pi/2$~pulse fields. One can then calculate $M_x$ after this sequence, which gives :
\begin{equation}
    \label{Mx}
M_x=\mathrm{Tr}\left[\rho_3\cdot S_x\right]=\frac{d}{2}\sin\omega_0\tau\sin\beta.
\end{equation}
One can easily see that in absence of free evolution, i.e. $\tau=0$,  $\mathrm{Tr}\left[\rho_3\cdot S_x\right]=0$, and there is no signal. This is due to the fact that in the $\tau=0$~case, the  $\pi/2\!-\!\tau\!-\!\pi/2$~sequence corresponds to a single $\pi$~pulse on the sample which indeed gives no signal. In fact, according to equation~(\ref{Mx}), providing that $\beta\neq 0$, $M_x\neq 0$ only if the nuclear spin state interference term $\sin\omega_0\tau$ is different from zero. In a $\pi/2-\tau-\pi/2$~sequence, the existence of any NMR signal is then the evidence of the occurrence of nuclear spin interferences.    
\section{Material and methods}
A sample of $10\ \mathrm{ml}$ of degassed water was placed at room temperature in a wide-bore magnet with a magnetic field of $4.7\ \mathrm{T}$ (Magnex). The NMR~spectrometer (SMIS) allows a phase precision of the RF~pulses of $0.25\ ~^{\circ}$. The RF~pulses had a gaussian shaped intensity with a duration $d=600\ \mu\mathrm{s}$, a frequency $\omega_0/2\pi=200,137\ \mathrm{MHz}$, and half-width of $3000\ \mathrm{Hz}$. The inter pulse delay between the ends of the first and second pulse was $\tau=1\ \mathrm{ms}$. The NMR~signal was detected in quadrature mode with a sample frequency of $5000\ \mathrm{Hz}$ and $8\ \mathrm{K}$~points. The intensity of the signal is obtained as the modulus of the $2$~parts given by the quadrature detection mode. The homogeneity of the magnetic field was measured by the line width obtained by Fourier Transform of the free induction decay (FID) acquired after a $\pi/2$~pulse. $\mathrm{T}1$ measured by inversion-recovery sequence was $3.2\ \mathrm{s}$ and $\mathrm{T}2$ measured by a Carr-Purcell-Meiboom-Gill sequence~\cite{cpmg} was $1.8\ \mathrm{s}$ slightly depending on the homogeneity of the magnetic field. NMR~spectrum of pure water, as for all liquid sample with no $J$~coupling, displays a very narrow line due to the motion averaging of the dipole-dipole coupling. Such a nuclear spin system is highly isolated from its surrounding and it is well-known that the relaxation time $\mathrm{T}1$  which characterizes the energy exchange with the lattice  and the inverse of the line width which measures the decoherence time are very long in high homogeneous magnetic field. 
\section{Results}
\begin{figure}
\includegraphics[scale=0.08]{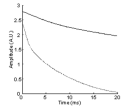}
\caption{NMR signal (FID) of $10\ \mathrm{ml}$ of water after one $\pi/2$~pulse. Continuous line is obtained in a high homogeneous magnetic field $(\Delta B_0/B_0= 2.0\cdot 10^{-8})$ and dashed line in less homogeneous field $(\Delta B_0/B_0= 2.7\cdot 10^{-7})$.}
\label{fig1A}
\end{figure}

\begin{figure}
\includegraphics[scale=0.8]{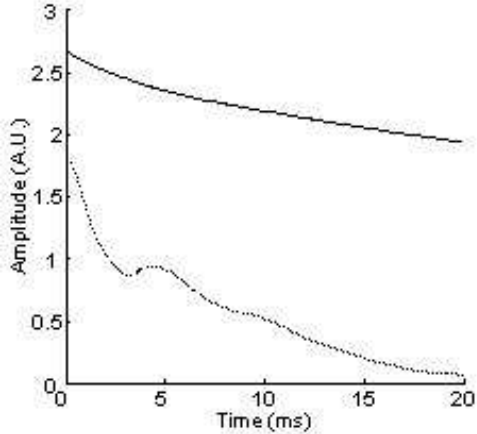}
\caption{NMR signal (FID) of $10\ \mathrm{ml}$ of water after two $\pi/2$~pulses with a relative phase of $\beta=90^{\circ}$. Continuous line is obtained in a high homogeneous magnetic field $(\Delta B_0/B_0= 2.0\cdot 10^{-8})$ and dashed line in less homogeneous field $(\Delta B_0/B_0= 2.7\cdot 10^{-7})$.}
\label{fig1B}
\end{figure}

\begin{figure}
\includegraphics[scale=0.25]{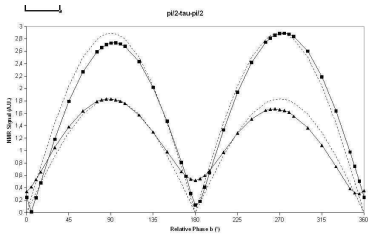}
\caption{Amplitude of the NMR~signal of $10\ \mathrm{ml}$ of water after two $\pi/2$~pulses versus the relative phase~$\beta$ of the two pulses. Continuous line ($\blacksquare\  \mathrm{NMR}_1$) is obtained in a high a homogeneous magnetic field, large dashed ($\blacktriangle\  \mathrm{NMR}_2$) line in a less homogeneous field. Fine dashed lines correspond to $f\left(\beta\right)=G\left|\sin\beta\right|$ normalized to the maximum NMR~signal in each case.}
\label{fig2}
\end{figure}

Experimentally, it was impossible for us to tune $\tau$ at a time scale small enough to vary $\omega_0\tau$ over $2\pi$. However, it is possible to ensure over typical experimental time (few minutes) an accurate stability of $\omega_0\tau$, i.e. the rms magnitude of the fluctuating part of this angle~$\omega_0\tau$ remains much smaller than $2\pi$. Under this last condition, one can then plot the NMR~signal given by the  $\pi/2\!-\!\tau\!-\!\pi/2(\beta)$~sequence as a function of $\beta$, the relative phase of the two $\pi/2$~pulse fields and compare the results to that given by equation~(\ref{Mx}). If the experimental data match equation~(\ref{Mx}), then the nuclear spin interference term is revealed and also controlled.\\
The NMR~signal (FID) after a single $\pi/2$~pulse is mainly dependent on the homogeneity of the magnetic field $B_0$. On Fig~\ref{fig1A}
one can see the FID recorded after a single $\pi/2$~pulse in a highly homogeneous field ($\Delta B_0/B_0= 2.0\cdot 10^{-8}$, continuous line) compared to a less homogeneous one ($\Delta B_0/B_0= 2.7\cdot 10^{-7}$ dashed line). With a $\pi/2\!-\!\tau\!-\!\pi/2$~sequence, it is  well known that NMR gives rise to an echo at time $t=\tau$ after the second $\pi/2$~pulse, described in~1950 by E.~Hahn as spin echo~\cite{hahn}. But, here we have measured the NMR~signal in a very homogeneous magnetic field and with small inter pulse delays where no spin echo is detected as seen on Fig.~\ref{fig1B} (continuous line).
Even in the less homogeneous magnetic field there is a modulation of the FID but no echo at $1\ \mathrm{ms}$ which is the delay between the  two $\pi/2$~pulses. The absence of echo in this case  is equivalent to the absence of any echo for an homogeneous line in an Electron Spin Resonance (ESR) experiment. The FID corresponds to the magnetization in the transverse plane and therefore the signal is proportional to $\sqrt{M_{x}^{2}+M_{y}^{2}}$. Fig.~\ref{fig2} shows the amplitude of the NMR signal at the beginning of the FID versus the relative phase~$\beta$.
As seen in Fig.~\ref{fig2}, in the case of a highly homogeneous magnetic field $(\Delta B_0/B_0= 2.0\cdot 10^{-8})$, $f\left(\beta\right)=G\left|\sin\beta\right|$ (given by equation~(\ref{Mx}) for a well defined value of $\omega_0\tau$) fits the experimental data pretty well. The maximum relative deviation $\Delta s\left(\beta\right)=\frac{\left(\mathrm{NMR}_1\left(\beta\right)
-f\left(\beta\right)\right)}{\max\left(\mathrm{NMR}_1\left(\beta\right)\right)}$ between the experimental curve $\mathrm{NMR}_1\left(\beta\right)$ and~$f(\beta)$ is found to be $\Delta s(15^{\circ})=9.7\%$. In the case of a less homogeneous field $(\Delta B_0/B_0= 2.7\cdot 10^{-7})$, the fit is less good and the maximum relative deviation is found to be $\Delta s(18^{\circ})=28.3\%$.

\section{Conclusion}
We have shown here, that if the homogeneity of the static and RF~magnetic fields are controlled at an enough high accuracy, we can control the nuclear spin interference term out the macroscopic sample of  water molecules up to a maximum relative deviation of $9.7\%$. This number may certainly be decreased by giving additional care to the experimental set up. However, these results are a first step to manipulation of nuclear spin coherence of water molecules or any appropriate liquid molecules with negligible interactions with their environment.\\
The main application of this macroscopic quantum behavior is to use magnetic gradients to define several qubits by their frequencies in a specific magnetic environment rather than qubits defined by the chemical environment in a molecule. Such an approach could lead into a more scalable~\cite{vinc} NMR~computer than the use of chemical molecules. If the reported experiments involve only one qubit, by using a magnetic gradient along the direction of the tube contenting the water sample, it could be defined a linear arrangement of several qubits. The same setup could be used to create and manipulate up to thirty qubits arranged along a line, thus realizing one-dimensional (and thus rather sophisticated but not universal~\cite{voor}) quantum cellular automata on $30$~qubits. We see no difficulties in extending our approach to two- and, possibly, three-dimensional settings implementing universal quantum cellular automata computers on at least as many qubits. Theses results on manipulation of nuclear spin coherence of water represent the first step in a radically new, scalable and easily reproducible approach to the field of quantum information processing based on liquid state NMR~techniques, defying in particular the recent skepticism of J.~A.~Jones~\cite{jones.a} on the viability of such techniques lately partially reversed by himself~\cite{jones.b} with the use of parahydrogen-derived compounds. 

\section{Acknowledgments}
J.~G. wishes to thank O.~Morizot for fruitful discussions. The assistance in the use of the $4.7\ \mathrm{T}$ NMR~spectrometer by T.~Guiberteau and the technical assistance of the $2\ \mathrm{T}$ NMR~spectrometer by C.~Marrer are acknowledged. E.~B. and D.~G. are grateful to Philippe Flajolet for his attention to the original project which has benefited from CNRS MathStic 2004 grant.

\end{document}